\documentclass[twocolumn]{aastex631}

\usepackage{comment}

\begin{document}

\title{Known changing-look AGN located within Rubin Deep Drilling Fields}

\correspondingauthor{Mariangella Camus}
\email{mariangella.camus@pucv.cl}

\author[0009-0000-1806-5302]{Mariangella Camus}
\affiliation{Pontificia Universidad Católica de Valparaíso, Chile}

\author[0000-0002-5854-7426]{Swayamtrupta Panda}\thanks{Gemini Science Fellow}
\affiliation{International Gemini Observatory/NSF NOIRLab, Casilla 603, La Serena, Chile}


\begin{abstract}

Changing-look active galactic nuclei (CL-AGN) exhibit spectroscopic and photometric changes on timescales of months to years, making them powerful laboratories for studying accretion variability onto supermassive black holes. Motivated by the growing relevance of large spectro-photometric time-domain surveys, especially the Vera C. Rubin Observatory LSST, we compiled a working literature-based catalog of known CL-AGN and evaluated its spatial overlap with the Rubin footprint. Using a geometric cross-match based on sky coordinates, we identify 79 sources located in the main survey area (Wide--Fast--Deep), including 5 sources lying within the high-cadence Deep Drilling Fields (DDFs). These DDF-associated sources are found in the COSMOS and XMM-LSS fields and represent promising targets for Rubin-era variability studies. The Rubin-overlap source list shown in this Research Note  is provided as a machine-readable table, including coordinates, field association, and literature reference information. A complete catalog release and broader multi-wavelength analysis will be presented in a forthcoming paper.

\end{abstract}

\keywords{Active galactic nuclei (16) -- Supermassive black holes (1663) -- Spectroscopy (1558) -- Time domain astronomy (2109) -- Optical observation (1169)
}

\section{Introduction} \label{sec:intro}

Changing-look active galactic nuclei (CL-AGN) are a remarkable class of accreting supermassive black holes that undergo substantial spectral and photometric transitions over relatively short timescales \citep{ ricci2023changing}. These changes may include the appearance or disappearance of broad emission lines, strong continuum variability, or transitions between different Seyfert-like spectral types \citep{lamassa2015discovery}. Because such behavior is thought to reflect major changes in the accretion flow, obscuration geometry, or both, CL-AGN provide a unique window into the physics of black hole feeding and the time-dependent structure of active galactic nuclei.\\

Over the last decade, the number of known CL-AGN has grown significantly thanks to repeated spectroscopic observations, long-baseline photometric monitoring, and dedicated searches in large surveys 
\citep{macleod2019changing, hon2022skymapper, guo2025changing}. However, the currently known population remains scattered across the literature, with different samples compiled using different selection criteria, naming conventions, and observational strategies. This fragmentation makes it difficult to carry out systematic time-domain studies, especially when trying to connect known CL-AGN with the next generation of synoptic surveys. A homogeneous compiled catalog is therefore a necessary step toward building a more complete observational framework for these objects.\\

This need is particularly relevant for connecting the currently known CL-AGN population with upcoming southern time-domain surveys. Several recent works have presented important CL-AGN samples or compilations \citep{amrutha2024discovering}, but these are necessarily limited by their selection methods, parent surveys, wavelength coverage, or spectroscopic follow-up strategy. Therefore, the list used here should not be interpreted as a complete census of all CL-AGN, but rather as a practical literature-based compilation designed to identify known sources that are already well placed for Rubin monitoring. Building such a list is useful for tracking previously reported CL-AGN, identifying objects suitable for future photometric and spectroscopic follow-up, and preparing for the classification of new Rubin-era candidates.\\


The Vera C. Rubin Observatory is going to transform this landscape through its unprecedented wide-field, multi-epoch view of the southern sky.\footnote{To date (15-June-2026), more than 11.57 million alerts have been processed from the Data Preview 1 survey. For more details, we refer readers to \url{https://lsst.fink-portal.org/stats}.}
 Even at this early stage, Rubin's observations already provide an opportunity to assess how known CL-AGN overlap with particularly high-cadence fields and, importantly, to help us organize time-domain strategies: (1) forecast variability via light-curve modeling, (2) detect new candidates by using established CL-AGN as templates, and (3) strategize follow-up observations (e.g., spectroscopy) for the most informative cases. In particular, the Rubin Deep Drilling Fields (DDFs) are especially attractive because they combine repeated observations with the strong multi-wavelength legacy of fields such as COSMOS and XMM-LSS \citep{Ivezic2019, 2026arXiv260121769P}. These regions, therefore, offer a natural starting point for assessing which known CL-AGN may already benefit from Rubin monitoring and which sources should be considered high-priority targets for future analysis.\\

 In this work, we present a spatial cross-match between this compiled list of known CL-AGN and Rubin observable footprint, with the specific goal of identifying targets located in Wide--Fast--Deep regions and within the Deep Drilling Fields. This Research Note is intended as a first-look study demonstrating the feasibility of linking known CL-AGN with Rubin survey geometry. The source list associated with the figure is made available with this Note, while the full, continuously updated catalog and multi-wavelength analysis will be presented separately.


\begin{figure*}
    \centering
    \includegraphics[width=\linewidth]{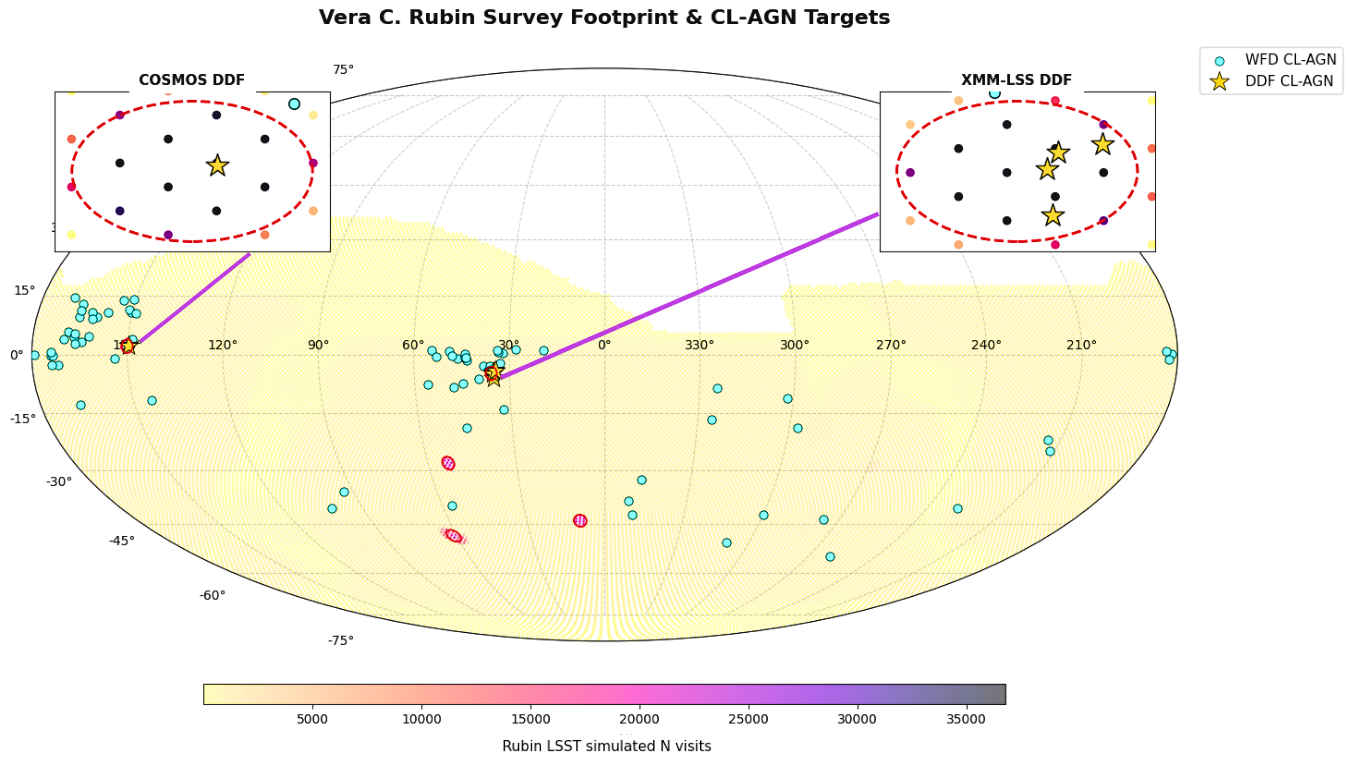}
    \caption{All-sky map showing the 79 compiled CL-AGN targets located within the adopted Rubin observing regions. The background shows the Rubin LSST cadence-simulation visit-count map. Cyan circles mark the 74 CL-AGN located in the Rubin Wide--Fast--Deep footprint, while gold stars mark the 5 CL-AGN located within the adopted Deep Drilling Field apertures. The inset panels show zoomed-in views of the two DDF regions containing CL-AGN in our compiled sample: COSMOS and XMM-LSS. Red dashed curves indicate the adopted DDF apertures, defined with radius $r=1.75^\circ$}
    \label{fig1}
\end{figure*}

\section{Analysis} \label{sec:analysis}
\subsection{Construction of the known CL-AGN catalog} \label{subsec:catalog}


We compiled a working list of known CL-AGN from the literature using the following source papers:
\citep{yang2018discovery,macleod2019changing,hon2022skymapper,temple2023bass,zeltyn2024exploring,panda2024changing,amrutha2024discovering,dong2025newly,guo2025changing,chen2026changing}.
These references constitute the complete set of literature sources used for the version of the compilation analyzed in this Research Note. The initial compilation included sources reported under different naming conventions and coordinate formats, requiring homogenization before any spatial analysis could be performed. Source names were standardized where possible, coordinates were expressed in a common format, and duplicate entries were removed through positional matching and manual inspection of ambiguous cases. This procedure yielded a clean working sample of 1,337 unique CL-AGN, which forms the basis of the present analysis. For transparency and reproducibility, the machine-readable table associated with this Note lists the Rubin-overlap sources shown in Fig.~\ref{fig1}, together with their coordinates, field association, and literature reference information. The full master catalog will be presented in a forthcoming paper.



\subsection{Rubin material used}
\label{subsec:rubin_material}


To evaluate the Rubin accessibility of the compiled CL-AGN sample, we compared the source positions with Rubin footprint and cadence information. We used the Rubin {\tt baseline\_v5.1.0\_10yrs} cadence simulation \citep[see e.g.,][]{2026arXiv260121769P} and selected high-cadence Wide--Fast--Deep regions using a visit-count threshold of $N \geq 804$ visits, corresponding to the high-cadence region shown in Fig.~\ref{fig1}. This criterion was applied uniformly across the footprint.\\

The Deep Drilling Fields were then treated as a separate geometric classification step, not as a different or more permissive matching procedure. In this work, we considered the Rubin DDF regions shown in Fig.~\ref{fig1}. We adopted circular apertures with radius $r=1.75^\circ$ around the DDF field centers. The corresponding on-sky area for each circular region is $A=\pi r^2 \approx 9.62~\mathrm{deg}^2$, which we use to define DDF membership in this work. Among the DDFs considered, only the COSMOS and XMM-LSS regions contain CL-AGN from our compiled sample, with centers at $(\alpha,\delta)=(150.12^\circ,+2.18^\circ)$ and $(35.71^\circ,-4.75^\circ)$ (J2000), respectively.

\subsection{Cross-match methodology}
\label{subsec:catalog}

The spatial comparison was performed using source sky positions and the Rubin footprint in a geometric cross-match framework. We first identified sources located in regions corresponding to the highest observational cadence, adopting a threshold based on the upper 20\% of visit counts. This selection isolates the most promising targets for future Rubin monitoring within the broader compiled CL-AGN population. We then applied a stricter spatial criterion to identify sources lying directly within the official DDF regions, and we verified the robustness of this selection by repeating the cross-match over a range of search radii, which yielded the same number of DDF-associated sources.\\

This two-step approach allows us to separate sources that are generally well placed within high-cadence Rubin areas from those that are especially valuable because they fall inside the best-monitored deep fields. Such a distinction is useful for prioritizing follow-up targets and for framing this note as a first practical bridge between the known CL-AGN population and Rubin time-domain data.\\

\section{Results and Discussions} \label{sec:results}

\subsection{Number of catalog sources and cross-match statistics}
\label{sec:results}

From the final sample of 1,337 compiled CL-AGN, we identify 79 sources that fall within the Rubin main survey footprint (Wide–Fast–Deep, WFD; see Fig.~\ref{fig1}). Among these, five sources are located within the Deep Drilling Fields (DDFs), while the remaining 74 lie in the broader WFD footprint outside the DDF regions. This result highlights that a non-negligible subset of known CL-AGN is already well-positioned for upcoming Rubin-based variability studies.

\subsection{Sources in specific fields}
\label{sec:results}

The five DDF-associated sources are distributed between the COSMOS and XMM-LSS fields, with one source in COSMOS and four sources in XMM-LSS. These fields are valuable because they combine high-cadence Rubin observations with extensive archival multi-wavelength coverage. CL-AGN located in these regions are therefore especially useful for future studies connecting Rubin variability with existing optical, infrared, and X-ray information.




\section{Summary and Future Work}
\label{sec:summary}

We compiled a working literature-based sample of 1,337 CL-AGN and performed a spatial cross-match with Rubin observing regions. This analysis identifies 79 sources in high-cadence Rubin Wide--Fast--Deep regions, including five sources located within the adopted DDF apertures, all in the COSMOS and XMM-LSS fields. These objects form a useful subset of known CL-AGN for future Rubin-based time-domain studies.\\

The Rubin-overlap source list shown in Fig.~\ref{fig1} is provided as a machine-readable table, including coordinates, field association, and literature reference information. This table is intended to make the present result reproducible while preserving the full master catalog and broader multi-wavelength analysis for a forthcoming paper. Future work will extend this effort by refining the CL-AGN compilation, incorporating additional literature and survey data, and analyzing Rubin light curves as further data products become available.

\begin{acknowledgments}
We thank the Scientific Editor for constructive comments that improved the clarity of this Research Note. MC is grateful for the hospitality and the staff at the International Gemini Observatory in La Serena, Chile. SP is supported by the International Gemini Observatory, a program of NSF NOIRLab, which is managed by the Association of Universities for Research in Astronomy (AURA) under a cooperative agreement with the U.S. National Science Foundation, on behalf of the Gemini partnership of Argentina, Brazil, Canada, Chile, the Republic of Korea, and the United States of America. We thank Matthew Temple (University of Durham, UK) for discussions on the BASS Changing-Look AGNs, and Arjun Dey (NOIRLab) for his help with accessing the DESI Legacy Imaging footprint.\\

\end{acknowledgments}

\vspace{5mm}
\facilities{ARC, ATLAS, ATT, Blanco, Bok, CTIO:1.5m, Du Pont, FTS, Hale, HET, ING:Herschel, KPNO:2.1m, LAMOST, LCOGT, Magellan:Baade, Magellan:Clay, Mayall, MMT, Perkins, PO:48, PS1, Skymapper, Sloan, UKST, VLT:Antu, VLT:Kueyen, WISE}

\software{numpy \citep{numpy}, matplotlib \citep{matplotlib}, astropy \citep{2013A&A...558A..33A,2018AJ....156..123A}}

\newpage
\bibliography{references}{}
\bibliographystyle{aasjournal}

\end{document}